\documentclass{article}
\begin{document}
\begin{center}
{\Large
{\bf Generalized Mittag-Leffler Distributions and Processes for Applications in Astrophysics and Time Series Modeling}}
\end{center}
\footnotesize
\begin{center}
\def\thefootnote{\fnsymbol{footnote}}
$Jose, K.K.^1 \footnote{Corresponding author},
Uma,P.^1, ~Seetha ~Lekshmi,V.^2, and ~Haubold,H.J.^3$\\
$^1$ Department of Statistics, St. Thomas College, Pala, M.G. University, Kerala-686 574. India.\\
$^2$ Department of Statistics, Nirmala College, Muvattupuzha, Kerala, India.\\
$^3$ Office for Outer Space Affairs, United Nations, P.O. Box 500, A-1400 Vienna, Austria.\\
\end{center}
\normalsize
\vskip.5cm

\noindent
{\bf Abstract.} Geometric generalized Mittag-Leffler distributions having the Laplace transform\\ 
$\frac{1}{1+\beta\log(1+t^\alpha)},0<\alpha\le 2,\beta>0$ is introduced and its properties are discussed. Autoregressive processes with Mittag-Leffler and geometric generalized Mittag-Leffler marginal distributions are developed. Haubold and Mathai (2000) derived a closed form representation of the fractional kinetic equation and thermonuclear function in terms of Mittag-Leffler function. Saxena et al (2002, 2004a,b) extended the result and derived the solutions of a number of fractional kinetic equations in terms of generalized Mittag-Leffler functions. These results are useful in explaining various fundamental laws of physics. Here we develop first-order autoregressive time series models and the properties are explored. The results have applications in various areas like astrophysics, space sciences, meteorology, financial modeling and reliability modeling.
\section{Introduction}
\normalsize

Recently Mittag-Leffler functions and distributions have received the attention of mathematicians, statisticians and scientists in physical and chemical sciences. Pillai (1990) introduced the Mittag-Leffler distribution in terms of Mittag-Leffler functions. Jayakumar and Pillai (1993) developed a first order autoregressive process with Mittag-Leffler marginal distribution. Fujita (1993) discussed a generalization of the results of Pillai. Jose and Seetha Lekshmi (1997) developed geometric exponential distribution and Seetha Lekshmi and Jose (2002, 2003) extended the results to obtain geometric Mittag-Leffler distributions. Jayakumar and Ajitha (2003) obtained various results on geometric Mittag-Leffler distributions. Seetha Lekshmi and Jose (2004) extend the concept to obtain geometric $\alpha$-Laplace processes.

A discrete version of the Mittag-Leffler distribution was introduced by Pillai and Jayakumar (1995). Lin (1998a,b, 2001) obtained various characterizations of the Mittag-Leffler distributions and refers to them as positive Linnik laws. In physics, Haubold and Mathai (2000) derived a closed form representation of the fractional kinetic equation and thermonuclear function in terms of Mittag-Leffler function. Saxena et al. (2004a,b) extended the result and derived the solutions of a number of fractional kinetic equations in terms of generalized Mittag-Leffler functions and obtained the solution of a unified form of generalized fractional kinetic equations, which provides the unification and extension of the earlier results. Such behaviors occur frequently in chemistry, thermodynamical and statistical analysis. In all such situations the solutions can be expressed in terms of generalized Mittag-Leffler functions. Weron and Kotulski (1986) use Mittag-Leffler distribution in explaining Cole-Cole relaxation.

 The function $E_\alpha(z)=\displaystyle\sum_{k=0}^{\infty}\left[\frac{z^k}{\Gamma(1+ak)}\right]$ was first introduced by Mittag-Leffler in 1903 (Erdelyi, 1955). Many properties of the function follow from Mittag-Leffler integral representation \\
\[E_\alpha(z)=\frac{1}{2\pi i}\int_{C}\frac{t^{\alpha-1}e^t}{t^\alpha-z}dt.\]
where the path of integration C is a loop which starts and ends at $-\infty$ and encircles the circular disc $|t|\le z^{\frac{1}{\alpha}}$. Pillai(1990) proved that $F_\alpha(x)= 1-E_\alpha(-x^\alpha),0<\alpha\le 1$ are distribution functions, having the Laplace transform $\psi(t)=(1+t^\alpha)^{-1}, t\ge 0$ which is completely monotone for $0<\alpha\le 1$. He called $F_\alpha (x)$, for $0<\alpha\le 1$, a Mittag-Leffler distribution. The Mittag-Leffler distribution is a generalization of the exponential distribution, since for $\alpha=1$, we get exponential distribution. Pillai (1990) has shown that $F_\alpha(x)$ is geometrically infinitely divisible (g.i.d.) and is in the domain of  attraction of stable laws.

Pillai (1985) developed $\alpha$-Laplace distribution with characteristic function given by  ${(1+|t|^\alpha)^{-1}}$; $0<\alpha \le 2$. This distribution is also known as Linnik distribution. Using a simple characterization of this distribution given by Linnik (1962), Anderson and Arnold (1993) constructed a discrete-time process having a stationary Linnik distribution. These Linnik models appear to be viable alternatives to stable processes as models for temporal changes in stock prices. Pakes (1998) gave a mixture representation for symmetric generalized Linnik laws as $A_{\alpha,v}\cong(\gamma(\nu))^{\frac{1}{\alpha}}S_\alpha$, where $A_{\alpha,v}$ denotes a random variable having the characteristic function $(1+|t|^\alpha)^{-v}, v>0$. This defines the generalized symmetric Linnik law $SLi(\alpha,v)$. Kotz and Ostrowski(1996) proved the following mixture representation of Linnik law in terms of another: If $0<\beta<\alpha\le 2$, then there is a random variable $V_{\alpha,\beta}(\nu)\ge0$ whose density is 
\[g(x,\alpha,\beta)=\frac{\alpha}{\pi}\sin(\frac{\pi\beta}{\alpha})\frac{x^{\beta-1}}{1+x^2\beta+2x^\beta\cos(\frac{\pi\beta}{\alpha})}\] such that $Y=XV_{\alpha,\beta}(\nu)$ where $X\stackrel{d}=AL(\alpha)$; $Y\stackrel{d}=AL(\beta)$. Jacques et al (1999) proved that the generalized Linnik laws belong to the Paretian family. They have shown that\\ $\displaystyle\lim_{n\rightarrow\infty}x^\alpha P[X>x]=\frac{\nu}{\pi}\Gamma(\alpha)\sin(\frac{\pi\alpha}{2})$, where X has the Linnik distribution. They also discussed the estimation of parameters of Linnik distribution. Kozubowski (2000) discussed the fractional moment estimation of Linnik parameters. Jayakumar et al. (1995) generalized the Laplace processes of Lawrance (1978) and Dewald and Lewis (1985).

Gaver and Lewis (1980) derived the exponential solution of first order autoregressive equation $X_n=\rho X_{n-1}+\eta_n, n=0,\pm1,\pm2,\ldots$, where $\{\eta_n\}$ is a sequence of independently and identically distributed random variables when $0\le \rho<1$. Lin (1994) proved the characterizations of the Laplace and related distributions via geometric compounding. It is well known that the concept of geometric compounding is related to the rarefaction of renewal processes (R$\acute{e}$nyi, 1956) and to damage models (Rao and Rubin, 1964; see also Galambos and Kotz, 1978, p.95).

In Section 2 of this paper, we describe some of the properties of generalized Mittag-Leffler distributions. In Section 3, we develop an autoregressive process of order one (AR (1)) with generalized Mittag-Leffler distribution. In Section 4, geometric generalized Mittag-Leffler distribution is introduced and their properties are discussed. Section 5 deals with autoregressive processes with the above marginal distribution. The extension of these results to $k^{th}$ order case is considered in \mbox{Section 6.} In Section 7 we describe some applications in various fields.
\section{Generalized Mittag-Leffler Distribution}
In this Section we introduce a new class of distributions called generalized Mittag-Leffler distribution denoted by GMLD ($\alpha,\beta$).\\
A random variable with support over $(0, \infty)$ is said to follow the generalized Mittag-Leffler distribution with parameters $\alpha$ and $\beta$ if its Laplace transform is given by \\
\begin{equation}
\psi(t)=E[e^{-tX}]=(1+t^\alpha)^{-\beta};0<\alpha \le 1,\beta>0.
\end{equation}
The cumulative distribution function (c.d.f.) corresponding to (1) is given by\\
\begin{equation}
F_{\alpha,\beta}(x)=P[X\le x]=\displaystyle\sum_{k=0}^\infty \frac {(-1)^k \Gamma (\beta+k)x^{\alpha(\beta+k)}}{k!\Gamma (\beta) \Gamma (1+\alpha(\beta+k))}
\end{equation}
It easily follows that when $\beta$=1, we get Pillai's Mittag-Leffler distribution (see Pillai, 1990).
When $\alpha$=1, we get the gamma distribution.
When $\alpha$=1, $\beta$=1 we get the exponential distribution.
This family may be regarded as the positive counterpart of Pakes generalized Linnik distribution characterized by the characteristic function 
\begin{equation}
\left(1+|t|^{\alpha}\right)^{-\beta};0<\alpha\le 2,\beta>0.
\end{equation}
(see Pakes, 1998).
Now we shall discuss some properties of generalized Mittag-Leffler distributions.
\vskip.3cm
\noindent
{\bf Theorem 2.1}\\ Let $U_\alpha$ follows the positive stable distribution with Laplace transform $\psi(t)=exp(-t^\alpha),\\ t>0; 0<\alpha\le 1.$ Let $V_\beta$ be a random variable, independent of $U_\alpha$, and follows a gamma distribution with parameter $\beta$ and Laplace transform $\phi(t)={\left(\frac{1}{1+t}\right)}^\beta; \beta>0.$ Then $X_{\alpha,\beta}=U_\alpha{V_\beta{^\frac{1}{\alpha}}}$ follows generalized Mittag-Leffler distribution GMLD($\alpha,\beta$).
\vskip.3cm
\noindent 
{\bf Proof:}\\
The Laplace transform of X is
\begin{eqnarray*}
\psi_X(t)&=&E(e^{-tVU^{\frac{1}{\alpha}}})\\
&=&\int_{0}^{\infty}\psi_U(tV^\frac{1}{\alpha})dF(v)\\
&=&\int_{0}^{\infty}e^{-t^\alpha V} dF(v)\\
&=&{\left(\frac{1}{1+t^\alpha}\right)}^\beta
\end{eqnarray*}
\vskip.3cm
\noindent 
{\bf Theorem 2.2}\\
The p.d.f. of $X_{\alpha,\beta}$ is a mixture of gamma densities.
\vskip.3cm
\noindent
{\bf Proof:}\\ 
Using Theorem 2.1, we have the c.d.f. of $X_{\alpha,\beta}$ as
\begin{equation}
F_{\alpha,\beta}(x)=\int_{0}^{\infty}S_{\alpha,v}(x)dF_\beta(v)
\end{equation}
where $S_{\alpha,v}(x)$ is the c.d.f. of a distribution with Laplace transform $exp(-vt^\alpha)$ and $F_\beta(v)$ is the c.d.f. of gamma distribution with p.d.f. $f(y)=\frac{1}{\Gamma \beta}y^{\beta-1}e^{-y}.$\\
We can rewrite (4) as
\[F_{\alpha,\beta}(x)=\int_{0}^{\infty} G_\beta (x/y)^\alpha dS_{\alpha,1}(y)\]
Hence the p.d.f. of X is 
\begin{eqnarray*}
f_{\alpha,\beta}(x)&=&\frac{d}{dx}F_{\alpha,\beta}(x)\\
&=& \int_{0}^{\infty}\frac{\alpha}{\Gamma \beta}\frac{x^{\alpha \beta-1}}{y^{\alpha \beta}}e^{-(x/y)^\alpha} dS_{\alpha,1}(y)
\end{eqnarray*}
This shows that $f_{\alpha,\beta}$ is a mixture of generalized gamma densities. For $\beta=1$, $ f_{\alpha,\beta}$ reduces to a mixture of Weibull densities.
\vskip.3cm
\noindent
{\bf Remark 2.1}\\
Lin (1998) has shown that $F_{\alpha,\beta}(x)$ is slowly varying at infinity, for $\alpha \in(0,1]$ and $\beta>0$.
\vskip.3cm
\noindent
{\bf Remark 2.2}\\
Pillai (1990) had obtained the fractional moments for $\beta=1$, as 
\[E(X_\alpha^r)=\frac{\Gamma (1-r/\alpha) \Gamma (1+r/\alpha)}{\Gamma (1-r)};0<r<\alpha \le 1.\]
In a similar manner Lin (1998) obtained the fractional moments for $X_{\alpha,\beta}$ for $0<\alpha\le 1$ and $\beta>0$ as
\begin{eqnarray*}
E(X_{\alpha,\beta}^r)&=&\frac{\Gamma (1-r/\alpha)\Gamma (\beta+r/\alpha)}{\Gamma (1-r) \Gamma \beta};-\alpha \beta<r<\alpha \\
&=& \infty \mbox{ if } r\le -\alpha \beta \mbox{ or } r\ge \alpha
\end{eqnarray*}
\vskip.3cm
\noindent 
{\bf Definition 2.1:} A probability distribution on $R=(0,\infty)$ is said to be in class L if its Laplace transform $\psi(t)$ satisfies
\begin{equation}
\psi(t)=\psi(kt)\psi_k(t), t\in R, k\in (0,1),
\end{equation}
with $\psi_k$ a Laplace transform.
\vskip.3cm
\noindent 
{\bf Theorem 2.3}\\
The GMLD belongs to the class L.
\vskip.3cm
\noindent 
{\bf Proof:}\\
The proof follows because of the relation 
\begin{equation}
(1+t^\alpha)^{-\beta}=(1+k^\alpha t^\alpha)^{-\beta}\left(k^\alpha+(1-k^\alpha)\frac{1}{1+t^\alpha}\right)^\beta
\end{equation}
{\bf Theorem 2.4}\\
The GMLD $(\alpha,\beta)$ is geometrically infinitely divisible for $0<\alpha\le 1, 0<\beta\le 1.$
\vskip.3cm
\noindent 
{\bf Proof:}\\
From Pillai and Sandhya (1990), a distribution is g.i.d. if and only if its Laplace transform is of the form $\psi(t)=\frac{1}{1+\phi(t)}$ where $\psi(t)$ has complete monotone derivative (c.m.d.) and $\phi(0)=0$. Now for GMLD ($\alpha, \beta)$, the Laplace transform is $\psi(t)=\frac{1}{1+\phi(t)}$ where $\phi(t)=(1+t^\alpha)^\beta-1.$ This has c.m.d. if and only if $0<\alpha\le 1$ and $0<\beta\le 1.$
\vskip.3cm
\noindent 
{\bf Remark 2.3}\\
Also being a mixture of gamma random variables, it is g.i.d. By Pillai and Sandhya (1990), GMLD  is a distribution with c.m.d. Hence it is infinitely divisible.
\section{First Order Autoregressive Processes with \\ GMLD$(\alpha,\beta)$ Marginals}
Now we shall construct a first order time series model with GMLD marginals.
The generalized Mittag-Leffler first order auto-regressive process GMLAR(1) is constituted by \{${X_n;n\ge1}$\} where $X_n$ satisfies the equation\\
\begin{equation}
 X_n = aX_{n-1}+ \eta_n ; 0<a<1,
\end{equation}
where $\{\eta_n\}$ is a sequence of independently and identically distributed random variables such that ${X_n}$ is stationary Markovian with generalized Mittag-Leffler marginal distribution. In terms of Laplace transform, eq. $(2)$ can be given as 
\begin{equation}
\psi_{X_n}(t)= \psi_{\eta_n}(t)\psi_{X_{n-1}}(at)
\end{equation}
Assuming stationarity, we have,\\
\begin{eqnarray}
\nonumber
\psi_{\eta }(t)&=&\frac{\psi_X(t)}{\psi_X(at)} \\
\nonumber
&=&\frac{(1+a^\alpha t^\alpha)^\beta}{(1+t^\alpha)^\beta}\\
\nonumber
&=&\left[\frac{1+{(at)}^\alpha}{1+t^\alpha}\right]^\beta\\
&=&\left[a^\alpha+(1-a^\alpha)\frac{1}{1+t^\alpha}\right]^\beta.
\end{eqnarray}
We can regard the innovations $\{\eta_n\}$ as the $\beta$-fold convolutions of random variables $U_n$'s such that 
\[U_n=\left\{\begin{array}{llll}
0& \mbox{with}& \mbox{probability}& a^\alpha\\ M_n& \mbox{with}&\mbox{probability}& 1-a^\alpha\end{array}\right.\]
where $M_n$'s are independently and identically distributed Mittag-Leffler random variables. 
Mittag-Leffler random variables can be generated easily using the following result. Let E be distributed as exponential with unit mean and let Q be distributed as positive stable with Laplace transform $e^{-t^\alpha}$ then $X=QE^{\frac{1}{\alpha}}$ will be distributed as Mittag-Leffler with Laplace transform $(1+t^\alpha)^{-1}$ (see Kozubowski and Rachev, 1999).

Jayakumar et al. (1995) developed an algorithm to generate Linnik random variables. In a similar manner, the GMLAR(1) process can be generated using computers.
If  $X_0\stackrel{d}=GMLAD(\alpha,\beta)$, then the process is strictly stationary. It is sufficient to verify that $X_n\stackrel{d}=GMLAD(\alpha,\beta)$ for every n. An inductive argument can be presented as follows. Suppose $X_{n-1}\stackrel{d}=GMLAD(\alpha,\beta)$.
Then from (8) we have,\\
\begin{eqnarray*}
\psi_X(t)= \left[\frac{1+{(at)}^\alpha}{1+t^\alpha}\right]^\beta \left[\frac{1}{1+{(at)}^\alpha}\right]^\beta =\left[\frac{1}{1+t^\alpha}\right]^\beta
\end{eqnarray*}
Hence the process is strictly stationary and Markovian provided $X_0$ is distributed as GMLD.
\vskip.3cm
\noindent 
{\bf Remark 3.1}\\
If $X_0$ is distributed arbitrarily, then also the process is asymptotically Markovian with generalized Mittag-Leffler marginal distribution.
\vskip.3cm
\noindent 
{\bf Proof:} 
\begin{eqnarray*}
X_n&=&aX_{n-1}+\eta_n\\
&=& a^nX_0+\displaystyle\sum_{k=0}^{n-1}{a^k\eta_{n-k}}.
\end{eqnarray*}
\noindent
Writing in terms of Laplace transform,
\begin{eqnarray*}
\psi_{X_n}(t)&=&\psi_{X_0}(a^nt)\displaystyle\prod_{k=0}^{n-1}\psi_\eta(a^kt)\\
&=&\psi_{X_0}(a^nt)\displaystyle\prod_{k=0}^{n-1}\left[\frac{1+{(a^{k+1}t)}^\alpha}{1+{(a^kt)^\alpha}}\right]^\beta \longrightarrow(1+t^\alpha)^{-\beta}  \mbox{as }  n \rightarrow \infty.
\end{eqnarray*}
Hence it follows that even if $X_0$ is arbitrarily distributed, the process is asymptotically stationary Markovian with generalized Mittag-Leffler marginals. We therefore have the following theorem.
\vskip.3cm
\noindent 
{\bf Theorem 3.1}\\
The first order autoregressive process $X_n=aX_{n-1}+\eta_n, a\in (0,1)$ is strictly stationary Markovian with generalized Mittag-Leffler marginal distribution as in (1) if and only if the $\{\eta_n\}$ are distributed independently and identically as the $\beta$-fold convolution of the random variable $\{U_n\}$ where\\ 
 \[U_n=\left\{\begin{array}{llll}
0& \mbox{with}& \mbox{probability}& a^\alpha\\ M_n& \mbox{with}&\mbox{probability}& 1-a^\alpha\end{array}\right.\]
where $\{M_n\}$ are independently and identically distributed Mittag-Leffler random variables provided $X_0\stackrel{d}=GMLD(\alpha,\beta)$ and independent of $\eta_n$. 
\vskip.3cm
\noindent 
{\bf Remark 3.2}\\
The model is defined for all values of 'a' such that $a \in (0,1)$. The autocorrelation is given by $\rho (r) =Cor(X_n,X_{n-r})=a^{\vert r\vert}$; $r=0,\pm 1,\pm 2,\ldots$ 
\subsection{Distribution of sums and bivariate distribution of $(X_n,X_{n+1})$}
We have,\\
\[X_{n+j}=a^jX_n+a^{j-1}\eta_{n+1}+a^{j-2}\eta_{n+2}+ \ldots+\eta_{n+j}; j=0,1,2,\ldots\]\\
Hence 
\begin{eqnarray*}
T_r&=&X_n+X_{n+1}+\ldots+X_{n+r-1}\\
&=&\displaystyle\sum_{j=0}^{r-1}[ a^jX_n+a^{j-1}\eta_{n+1}+a^{j-2}\eta_{n+2}+ \ldots+\eta_{n+j}]\\
&=&X_n\left[\frac{1-a^r}{1-a}\right]+\displaystyle\sum_{j=1}^{r-1}\eta_{n+j}\left[\frac{1-a^{r-j}}{1-a}\right]. 
\end{eqnarray*}
Therefore the distribution of the sums $T_r$ is uniquely determined by the Laplace transform
\begin{eqnarray*}
\psi_{T_r}(t)&=&
\psi_{X_n}\left(\frac{1-a^r}{1-a}t\right)\displaystyle\prod_{j=1}^{r-1}\psi_\eta\left(\frac{1-a^{r-j}}{1-a}t\right)\\
&=& \frac{1}{\left[1+\left(\frac{1-a^r}{1-a}t\right)^\alpha\right]^\beta}\displaystyle\prod_{j=1}^{r-1}\left[a^\alpha+(1-a)^\alpha\frac{1}{\left[1+\left(\frac{1-a^{r-j}}{1-a}t\right)^\alpha\right]^\beta}\right].
\end{eqnarray*}
The distribution of $T_r$ can be obtained by inverting the above expression. Next, the joint distribution of contiguous observations $(X_n,X_{n+1})$ can be given in terms of bivariate Laplace transform as,
\begin{eqnarray*}
\psi_{X_n,X_{n+1}}(t_1,t_2)&=&E[exp(it_1X_n+it_2X_{n+1})]\\
&=& E[exp(it_1X_n+it_2(aX_n+\eta_n))]\\
&=& E[exp(i(t_1+at_2)X_n+it_2\eta_{n+1})]\\
&=&\psi_{\eta_n}(t_2)\psi_{X_n}(t_1+at_2)\\
&=&\left[\frac{1+(at_2)^\alpha}{1+t_2^\alpha}\right]^\beta \left[\frac{1}{1+(t_1+at_2)^\alpha}\right]^\beta.
\end{eqnarray*}
Since this expression is not symmetric in $t_1$ and $t_2$, it follows that the GMLAR(1) process is not time reversible.
\section{Geometric Generalized Mittag-Leffler \mbox{distribution}}
Geometric generalized Mittag-Leffler distribution is introduced and some of its properties are studied.
\vskip.3cm
\noindent 
{\bf Definition 4.1}: A random variable X on $R=(0,\infty)$ is said to follow geometric generalized Mittag-Leffler distribution and write $X\stackrel{d}=GGMLD(\alpha,\beta)$ if it has the Laplace transform
\begin{eqnarray}
\psi(t)&=& \frac{1}{1+\beta \mbox{log}(1+t^\alpha)}, 0<\alpha \le2,\beta>0.
\end{eqnarray}
\vskip.3cm
\noindent 
{\bf Remark 4.1}\\ Geometric Generalized Mittag-Leffler distribution is geometrically infinitely divisible.
\vskip.3cm
\noindent 
{\bf Theorem 4.1}\\
Let $X_1,X_2,\ldots$ are independently and identically distributed Mittag-Leffler random variables and $N(p)$ be geometric with mean $\frac{1}{p}$, $P[N(p)=k]$=$p(1-p)^{k-1}$, $k=1,2,\ldots,0<p<1$. Define $Y=X_1+X_2+\ldots+X_{N(p)}$, then $Y\stackrel{d}=GGMLD(\alpha,\beta)$.
\vskip.3cm
\noindent 
{\bf Proof:}\\
The Laplace transform of Y is 
\begin{eqnarray*}
\psi_Y(t)&=&\displaystyle\sum_{k=1}^{\infty}[\psi_X(t)]^kp(1-p)^{k-1}\\
&=& \frac{1}{1+\frac{1}{p}\mbox {log}(1+t^\alpha)}.
\end{eqnarray*}
Hence $Y\stackrel{d}=GGMLD(\alpha,\frac{1}{p})$.
\vskip.3cm
\noindent 
{\bf Theorem 4.2}\\
Geometric generalized Mittag-Leffler distribution is the limit of geometric sum of GML$(\alpha,\frac{\beta}{n})$ random variables.
\vskip.3cm
\noindent 
{\bf Proof:}\\
$(1+t^\alpha)^{-\beta}= \{1+(1+t^\alpha)^{\frac{\beta}{n}}-1\}^{-n}$ is the Laplace transform of a probability distribution since generalized Mittag-Leffler distribution is infinitely divisible. Hence by lemma 3.2 of Pillai (1990),
\[\psi_n(t)=\{1+n[(1+t^\alpha)^{\frac{\beta}{n}}-1]\}^{-n}\],
is the Laplace transform of a geometric sum of independently and identically distribute generalized Mittag-Leffler random variables. Taking limit as $n\rightarrow \infty$\\
\begin{eqnarray*}
\psi(t)&=& \displaystyle\lim_{n\rightarrow\infty}\psi_n(t)\\
&=&\{1+\displaystyle\lim_{n\rightarrow\infty}n[(1+t^\alpha)^{\frac{\beta}{n}}-1]\}^{-1}\\
&=&[1+\beta\mbox{log}(1+t^\alpha)]^{-1}.
\end{eqnarray*}
\vskip.3cm
\noindent 
{\bf Theorem 4.3}\\
If W and V are independent random variables such that W has geometric gamma distribution with Laplace transform $\frac{1}{1+\beta\mbox{log}(1+t)}$ and V has a  positive stable distribution having Laplace transform $e^{{- t}^\alpha}$, then $W^{\frac{1}{\alpha}}V=U$ where $U\stackrel{d}=GGMLD(\alpha,\beta).$
\vskip.3cm
\noindent 
{\bf Proof:}\\
The Laplace transform of U is  
\begin{eqnarray*}
\psi_u(t)&=&E(e^{-tW^{\frac{1}{\alpha}}V})\\
&=&\int_{0}^{\infty}\psi_V(tW^\frac{1}{\alpha})dF(w)\\
&=&\int_{0}^{\infty}e^{-Wt^\alpha} dF(w)
=\frac{1}{1+\beta\mbox{log}(1+t^\alpha)}.
\end{eqnarray*}
\section{Geometric Generalized Mittag-Leffler Processes}
In this section, we develop a first order new autoregressive process with geometric generalized Mittag-Leffler marginals.\\
Consider an autoregressive structure given by,
\begin{equation}
X_n=\left\{\begin{array}{llll}
\eta_n&\mbox{with}&\mbox{probability}&p\\
X_{n-1}+\eta_n&\mbox{with}&\mbox{probability}&(1-p)
\end{array}\right.
\end{equation}
where $0<p<1$. Now we shall construct an AR (1) process with stationary marginal distribution as geometric generalized Mittag-Leffler distribution $GGMLD (\alpha,\beta)$.
\vskip.3cm
\noindent 
{\bf Theorem 5.1}\\
Consider a stationary autoregressive process $\{X_n\}$ with structure given by (11). A necessary and sufficient condition that $\{X_n\}$ is stationary Markovian with geometric generalized Mittag-Leffler marginal distribution is that $\{\eta_n\}$ is distributed as geometric Mittag-Leffler provided $X_0$ is distributed as geometric generalized Mittag-Leffler.
\vskip.3cm
\noindent 
{\bf Proof:}\\
Let us denote the Laplace transform of $X_n$ by $\psi_{X_n}(t)$ and that of $\eta_n$ by $\psi_{\eta_n}(t)$, equation (11) in terms of Laplace transform becomes\\
\[\psi_{X_n}(t)=p\psi_{\eta_n}(t) + (1-p) \psi_{X_{n-1}}(t) \psi_{\eta_n}(t)\]. 
On assuming stationarity, it reduces to the form
\begin{eqnarray*}
\psi_X(t)&=&p\psi_{\eta}(t) + (1-p) \psi_X(t)\psi_{\eta}(t).\\
\noalign{\noindent\mbox {Writing}}\psi_X(t)&=&\frac{1}{1+\beta\mbox{log}(1+t^\alpha)}~~~~\mbox
{and solving we get},\\
\psi_{\eta}(t)&=&\frac{1}{1+\beta\mbox{p log}(1+t^\alpha)}
\end{eqnarray*}
Hence it follows that $\eta_n\stackrel{d}=GGMLD(\alpha,p\beta)$.\\
The converse can be proved by the method of mathematical induction as follows. Now assume that $X_{n-1}\stackrel{d}=GGMLD(\alpha,\beta)$.
Then 
\begin{eqnarray*}
\psi_{X_{n-1}}(t)&=&\psi_{\eta_n}(t)\left[p+(1-p)\psi_{X_{n-2}}(t)\right]\\
&=&\frac{1}{1+p\beta\log(1+t^\alpha)}\left[p+(1-p)\frac{1}{1+\beta\log(1+t^\alpha)}\right]\\
&=&{[1+\beta \log(1+t^\alpha)]}^{-1}.
\end{eqnarray*}
\vskip.3cm
\noindent 
{\bf Remark 5.1}\\
Note that $X_n$ and $\eta_n$ belongs to the same family of distributions.

\subsection{The joint distribution of $X_n$ and $X_{n-1}$}
Consider the autoregressive structure given in (11). It can be rewritten as
\begin{eqnarray*}
X_n&=&I_nX_{n-1}+\eta_n \mbox,\\
where\\
P[I_n=0]&=&1-P[I_n=1]=p, 0<p<1.
\end{eqnarray*}
Then the joint Laplace transform of $(X_n,X_{n-1})$ is given by,
\begin{eqnarray*}
\psi_{X_n,X_{n-1}}(t_1,t_2)&=&E(e^{it_1X_{n-1}+it_2X_n)}\\
&=&E(e^{it_1X_{n-1}+it_2(I_nX_{n-1}+\eta_n})\\
&=&E(e^{(it_1+it_2I_n)X_{n-1}})\psi_{\eta_n}(t_2)\\
&=&\frac{1}{1+\beta\mbox{p}\log(1+t_2^\alpha)}\left[\frac{p}{1+\beta\log(1+t_1^\alpha)}+\frac{1-p}{1+\beta\log(1+{t_1+t_2}^\alpha)}\right].
\end{eqnarray*}
This shows that the process is not time reversible.
\section{Generalization to a $k^{th}$ order geometric generalized Mittag-Leffler Autoregressive processes}
Lawrance and Lewis (1982) constructed higher order analogs of the autoregressive eq. (11) with structure as given below,
\begin{equation}
X_n=\left\{\begin{array}{llll}
\eta_n&\mbox{with}&\mbox{probability}&p\\
X_{n-1}+\eta_n&\mbox{with}&\mbox{probability}&p_1\\
X_{n-2}+\eta_n&\mbox{with}&\mbox{probability}&p_2\\
\vdots\\
X_{n-k}+\eta_n&\mbox{with}&\mbox{probability}&p_k
\end{array}\right.
\end{equation}
where $p_1+p_2+\ldots+p_k=1-p , 0\le p_i,p\le 1, I=1,2,\ldots,k$ and $\eta_n$ is independent of  $\{X_n,X_{n-1},\ldots\}$.\\
In terms of Laplace transform, eq. (7) can be given as 
\[ \psi_{X_n}(t)= p\psi_{\eta_n}(t)+p_1\psi_{X_{n-1}}(t) \psi_{\eta_n}(t)+ p_2\psi_{X_{n-2}}(t) \psi_{\eta_n}(t)+\ldots+ p_k\psi_{X_{n-k}}(t) \psi_{\eta_n}(t)\]
Assuming stationarity, we get,\
\[\psi_{\eta_n}(t)= \frac{\psi_X(t)}{p+(1-p)\psi_X(t)}\].
This establishes that the results developed in Section 5 are valid in this case also. This gives rise to the $k^{th}$ order geometric generalized Mittag-Leffler autoregressive processes.
\section{Applications}
In thermodynamical or statistical applications, one is interested in mean values of a quantity Z(t). Tsallis (1988) generalized  the entropic functional of Boltzmann-Gibbs statistical mechanics that leads to q-exponential distributions. He used the mathematical simplicity of kinetic-type equations to emphasize the natural outcome of this distribution that corresponds exactly to the solution of the kinetic equation of non-linear type; the solution has power-law behavior. Saxena et al. (2004) showed that the fractional generalization of the linear kinetic-type equation also leads to power-law behavior. In both cases, solutions can be expressed in terms of generalized Mittag-Leffler functions. Mittag-Leffler distributions can also be used as waiting-time distributions as well as first-passage time distributions for certain renewal processes. Pillai (1990) developed renewal processes with geometric exponential as waiting time distribution. In a similar manner renewal processes with generalized Mittag-Leffler and geometric generalized Mittag-Leffler waiting times can be constructed.
\vskip.3cm
\noindent
{\bf Acknowledgment}
\vskip.3cm
\noindent
The second author is grateful to Kerala State Council for Science, Technology and Environment for the KSCSTE Fellowship under which this research was conducted.
\begin{center}
{\large
\noindent
{\bf References}}
\end{center}
\begin{enumerate}
\item
Anderson, D.N., Arnold, B.C.: Linnik distributions and processes. Journal of Applied Probability {\bf 30}, 330-340 (1993)

\item
Dewald, D.N., Lewis, P.A.W.: A new Laplace second order auto regressive time series model-NLAR (2). IEEE Transactions in Information Theory {\bf 31}, 645-651 (1985) 

\item
Devroye, L.: A note on Linnik's distribution. Statistical Probability Letters {\bf 9}, 305-306 (1990)

\item
Erd$\acute{e}$lyi, A.: Higher Transcendental Functions. Vol. 3, McGraw Hill, New York (1955)

\item
Fujita, Y.: A generalization of the results of Pillai. Annals of the Institute of Statistical Mathematics {\bf 45} 361-365 (1993)

\item
Galambos, J., Kotz, S.: Characterizations of probability distributions. Lecture Notes in Mathematics, Vol. 675, Springer-Verlag, New York (1978) 

\item
Gaver, D.P., Lewis, P.A.W.: First-order autoregressive gamma sequences and point processes. Advances in Applied Probability {\bf 12} 727-745 (1980)

\item
Haubold, H.J., Mathai,A.M.: The fractional kinetic equation and thermonuclear functions. Astrophysics and Space Science {\bf 273} 53-63 (2000)

\item
Jayakumar, K., Pillai, R.N.: On class L distributions. Journal of the Indian Statistical Association {\bf 30} 103-108 (1992)

\item
Jayakumar, K., Pillai, R.N.: The first-order autoregressive Mittag-Leffler process. Journal of Applied Probability {\bf 30} 462-466 (1993)

\item
Jayakumar, K., Kalyanaraman, K., Pillai, R.N.: $\alpha$-Laplace Processes. Mathematics of Computational Modelling {\bf 22} 109-116 (1995)

\item
Jayakumar,K., Ajitha, B.K.: On the geometric Mittag-Leffler distributions. Calcutta Statistical Association Bulletin {\bf 54}, Nos. 215-216, 195-208 (2003)

\item
Jose, K.K., Seetha Lekshmi, V.: On geometric exponential distribution and its applications. Journal of the Indian Statistical Association {\bf 37}, 51-58 (1997).

\item
Klebanov, L.B., Maniya, G.M., Melamed, I.A.: A problem of Zolotarev and analogs of infinitely divisible and stable distributions in a scheme for summing a random number of random variables. Theory of Probability Applications {\bf 24} 791-794 (1984)

\item
Kotz, S., Kozubowski, T.J., Podgorski, K.: The Laplace Distributions and Generalizations. Birkhaeuser, Boston (2001)

\item
Kotz, S., Ostrovskii, I.V.: A mixture representation of the Linnik distribution. Statistical Probability Letters {\bf 26}, 61-64 (1980)

\item
Kozubowski, T.J., Rachev, S.T.: Univariate geometric stable laws. Journal of Computational Analysis and Applications, preprint (1999)

\item
Kozubowski, T.J.: Fractional moment estimation of Linnik and Mittag-Leffler parameters. Mathematics of Computational Modelling. Special issue: Stable Non-Gaussian models in finance and econometrics. {\bf 34}, 1023-1035 (2000)

\item
Kozubowski, T.J.: Mixture representation of Linnik distribution revisited. Statistical Probability Letters {\bf 38}, 157-160 (1998)

\item
Lawrance, A.J.: Some autoregressive models for point processes. Colloquia Mathematica Societatis Janos Bolyai {\bf 24} Point Processes and Queuing Problems, Hungary, 257-275 (1978)

\item
Lin, G.D.: A note on the characterization of positive Linnik laws. Australian New Zealand Journal of Statistics {\bf 43} 17-20 (2001)

\item
Lin, G.D.: A note on the Linnik distributions. Journal of Mathematical Analysis and Applications {\bf 217} 701-706 (1998a)

\item
Lin, G.D.: On the Mittag-Leffler distributions. Journal of Statistical Planning Inference {\bf 74} 1-9 (1998b)

\item
Lin, G.D.: Characterizations of the Laplace and related distributions via geometric compound. Sankhya {\bf 56} 1-9 (1994)

\item
Linnik, Yu. V.: Linear forms and statistical criteria, I, II. Ukrainian Mathematical Zhurnal {\bf 5} 207-243 ((1962), English Translations in Mathematical Statistics and Probability {\bf 3}, 1-40, 41-90, American Mathematical Society, Providence, R.I.

\item
Pakes, A.G.: Mixture representations for symmetric generalized Linnik laws. Statistical Probability Letters {\bf 37} 213-221 (1998)

\item
Pillai, R.N.: Semi-$\alpha$-Laplace distributions. Communications in Statistical Theoretical Methods {\bf 14} 991-1000 (1985)

\item
Pillai, R.N.: On Mittag-Leffler and related distributions. Annals of the Institute of Statistical Mathematics {\bf 42} 157-161 (1990)

\item
Pillai, R.N., Sandhya, E.: Distributions with complete monotone derivative and geometric infinite divisibility. Advances in Applied Probability {\bf 22} 751-754 (1990)

\item
Pillai, R.N., Jayakumar, K.: Specialized class L property and stationary autoregressive process. Statistical Probability Letters {\bf 19} 51-56 (1994)

\item
Pillai, R.N., Jayakumar, K.: Discrete Mittag-Leffler distributions. Statistical Probability Letters {\bf 23} 271-274 (1995)

\item
{\bf Rao, C.R.} and {\bf Rubin,H. }(1964). On characterization of the Poisson distribution, {\it Sankhya}, Ser.A, {\bf 26}, 294-298.

\item
R$\acute{e}$nyi, A.: A characterization of the Poisson process. Magyar Tud. Akad. Mat. Kutato Int. Kozl. {\bf 1}, 519-527 (1956) (in Hungarian).(Translated into English in Selected Papers of Alfred R$\acute{e}$nyi, Vol.1, Akademiai Kiad$\acute{o}$, Budapest, 1976)

\item
Saxena, R.K., Mathai, A.M., Haubold, H.J.: Unified fractional kinetic equation and a fractional diffusion equation. Astrophysics and Space Science {\bf 209} 299-310 (2004a)

\item 
Saxena, R.K., Mathai, A.M., Haubold, H.J.: On generalized fractional kinetic equations. Physica A {\bf 344} 657-664 (2004b)

\item
Seetha Lekshmi, V., Jose, K.K.: Geometric Mittag-Leffler tailed autoregressive processes. Far East Journal of Theoretical Statistics {\bf 6} 147-153 (2002)

\item
Seetha Lekshmi, V., Jose, K.K.: Geometric Mittag-Leffler distributions and processes. Journal of Applied Statistical Sciences (accepted for publication) (2003)

\item 
Seetha Lekshmi, V., Jose, K.K.: An autoregressive process with geometric $\alpha$-Laplace marginals. Statistical Papers {\bf 45} 337-350 (2004)

\item
Tsallis, C.: Possible generalization of Boltzmann-Gibbs statistics. Journal of Statistical Physics {\bf 52} 479-487 (1988)

\item
Weron, K., Kotulski, M.: On the Cole-Cole relaxation function and related Mittag-Leffler distribution. Physica A {\bf 232} 180-188 (1996)
\end{enumerate}
\end{document}